\documentclass[twocolumn,showpacs,preprintnumbers,amsmath,amssymb]{revtex4}
\usepackage{amsfonts}

\usepackage{graphicx}% Include figure files
\usepackage{dcolumn}% Align table columns on decimal point
\usepackage{bm}% bold math
\newcommand{\sss}{\scriptscriptstyle}

\newcommand {\be}{\begin{equation}} % start equation
\newcommand{\ee}{\end{equation}}    % end equation

\def\dds1{\frac{\partial}{\partial s_1}}

\def\vte{v_{{\sss T}e}}

\def\d{d\kern-0.8 ex\vrule height 1.3 ex depth-1.24 ex width 0.7 ex
\kern 0.15 ex}
\def\D{D\kern-1.7 ex\vrule height .87 ex depth-0.8 ex width 0.7 ex
\kern 0.95 ex}

\def\bq{\begin{equation}}
\def\eq{\end{equation}}

\begin{document}

\preprint{PRE}

\title{Drift wave stabilized by an additional streaming ion or plasma population  }

\author{M. F. Bashir${}^{1}$, J. Vranjes${}^{2, 3}$ }
 \affiliation{
$^{1}$Department of Physics, COMSATS Institute of Information Technology, Lahore
54000, Pakistan\\
$^{2}$Instituto de Astrofisica de Canarias, 38205 La Laguna, Tenerife, Spain\\
$^{3}$Departamento de Astrofisica, Universidad de La Laguna, 38205 La Laguna, Tenerife, Spain.}

\date{\today}

\begin{abstract}
It is shown  that the universally unstable kinetic drift wave
in an electron-ion plasma can very effectively be suppressed by adding an extra flowing  ion (or plasma) population.   The effect of the flow of the added  ions is essential, their response is of the type $(v_{ph}-v_{f0})\exp[-(v_{ph}-v_{f0})^2]$, where $v_{f0}$ is the flow speed and $v_{ph}$ phase speed parallel to the magnetic field vector. The damping is strong and it is mainly due to this ion exponential term, and this remains so for $v_{f0}<v_{ph}$.
\end{abstract}

\pacs{52.35.Kt, 52.30.Gz, 52.35.Qz}
%\keywords{Pair-ion plasmas, Backward waves, Ion cyclotron resonances}

\maketitle

\section{Introduction}

Drift wave is called universally growing mode due to the fact that it is unstable
in both fluid and kinetic descriptions, and in collisional and collisionless plasmas. The wave is self-excited and it grows
due to the free energy in  plasma inhomogeneity and this remains so even in plasmas with hot ions. This fact
was used in our recent papers where a new paradigm was put forward for the heating of the solar corona  \cite{v1, v2} and solar wind
\cite{v3} by drift waves, based on stochastic heating mechanism from Refs.~\cite{mc, san}.
In practical situations in lab plasmas, in order to study the mode in  a controlled
situation, the wave   is on purpose excited and driven by an  electron current \cite{hat} or by a shear flow \cite{kan}. More recent experimental studies of drift instabilities are available
 in Refs.~\cite{xa1, xa2}.

 This is a dangerous mode in any  plasma environment  \cite{vlad}-\cite{na2}, and various effects have been studied in the past in order to stabilize it.
  One of them is the magnetic shear which in simple slab geometry introduces a layer, in the direction of the shear gradient, at which the mode is stabilized by resonant ions \cite{per, wei}. So although the stabilization is kinetic by nature it is routinely described as an effect of the plasma geometry.

 But in more  realistic laboratory situations, the same geometry which implies the stabilization by the magnetic shear in fact includes some additional features, like toroidal mode coupling, which may completely cancel the magnetic shear stabilization \cite{tay}. Much more on these phenomena may be found in our earlier works \cite{ps93, pop94, ps95}.

 Yet another way of the drift wave stabilization is by  cold electrons added to the plasma.  In the present work we show that this can also be done by adding flowing {\em ions or plasma}, which  need not be cold at all.

\section{The model and derivations}

The geometry assumed in the derivation is such that the
background magnetic field ($\mathbf{B}_{0}$) is oriented along the $z$-axis.  We assume a static (denoted by index $s$) inhomogeneous and quasineutral electron-ion plasma $n_{es0}(x)=n_{is0}(x)$, penetrated by a homogeneous  plasma stream (index $f$ further in the text). We allow for the presence of electrons as well in the $f$-species in order to  avoid the issue of excess charge in case that ions alone are added, i.e., $n_{ef0}=n_{if0}=const.$, although stabilization is mainly by the $f$-ions.  The equilibrium density gradient is in the $x$-direction and the wave vector $%
\mathbf{k}$ lies in the $y, z$-plane.

Electrostatic perturbations are assumed propagating  nearly perpendicular to the magnetic field   $\sim \exp(-i \omega t + i\mathbf{ k_{\bot }}\mathbf{r} + i k_z z)$.
 The perturbed densities are:
\be
\frac{n_{es1}}{n_{es0}}\!=\!\frac{e\phi _{1}}{T_{se}}\!\left\{\! 1\!+\!\left(\omega -\omega
_{se}^{\ast }\right)\!\! \sum_{n=-\infty }^{\infty }\!\!\!\left[ W(\xi
_{nse})\!-1\!\right]  \frac{\Lambda _{n}(b_{se})}{\omega\! +\!n\Omega _{se}}\! \right\}\!,  \label{3}
\ee
\be
\frac{n_{is1}}{n_{is0}}\!=\!-\frac{e\phi _{1}}{T_{si}}\!\left\{ \!1\!+\!\left( \omega -\omega
_{si}^{\ast }\right)\!\! \!\sum_{n=-\infty }^{\infty }\!\!\left[\! W(\xi
_{nsi})\!-1\!\right]\!  \frac{\Lambda _{n}(b_{si})}{\omega \!-\!n\Omega _{si}}%
\!\right\},   \label{4}
\ee
\be
\frac{n_{ef1}}{n_{ef0}}=\frac{e\phi _{1}}{T_{fe}}\left\{ 1+\widetilde{\omega }%
\!\!\sum_{n=-\infty }^{\infty }\!\!\left[ W(\widetilde{\xi }_{fe})-1\right] \right.
 \left. \frac{\Lambda _{n}(b_{fe})}{\widetilde{\omega }+n\Omega _{fe}}%
\right\},
\label{5}
\ee
\begin{equation}
\frac{n_{if1}}{n_{if0}}\!=\!-\frac{e\phi _{1}}{T_{fi}}\left\{ 1+\widetilde{\omega }%
\!\!\sum_{n=-\infty }^{\infty }\!\!\left[ W(\widetilde{\xi }_{fi})-1\right] \right.
\left. \frac{\Lambda _{n}(b_{fi})}{\widetilde{\omega }-n\Omega _{fi}}\right\}.  \label{6}
\end{equation}%
Here,
\[
W(\widetilde{\xi }_{n(s,f)\alpha })=\sqrt{2\pi }\int\limits_{-\infty
}^{\infty }\frac{xe^{-x^{2}/2}}{x-\widetilde{\xi }_{n(s,f)\alpha }}dx,
\vspace{-0.3cm}
\]%
and
\[
\widetilde{\xi }_{n(s,f)\alpha }=\frac{\widetilde{\omega }-n\Omega
_{(s,f)\alpha }}{k_zv_{t(s,f)\alpha }}, \quad  b_{(s,f)\alpha }=\frac{k_{\bot }^{2}v_{t(s,f)\alpha }^{2}}{\Omega _{(s,f)\alpha }},
\]
\[
\omega _{s\alpha }^{\ast }=-\frac{k_{\bot }v_{t\alpha }^{2}}{\Omega_{s\alpha }L_{ns\alpha }}, \quad \Omega_{(s,f)\alpha }=\frac{q_{(s,f)\alpha}B_{0}}{m_{(s,f)\alpha }},
\]
$\alpha $ denotes the species, $q_{\alpha }$ is their charge, $n_{\alpha 0 }$ is the equilibrium density, $L_{ns\alpha }$ is the inhomogeneity scale length of static component and $\widetilde{\omega }=\omega -k_zv_{\alpha 0 }$ is the Doppler shifted frequency due to the streaming velocity $v_{\alpha 0 }$.

 We are considering the
case of low plasma beta $\beta _{\alpha }=2\mu _{0}n_{\alpha 0}T_{\alpha
}/B_{0}^{2}<<1$ due to which the magnetic field gradient is ignored
following the relation $L_{ns\alpha }/L_{Bs\alpha }\sim \beta _{\alpha }$ \cite{v3},
where $L_{Bs\alpha }$  is the scale length of magnetic field inhomogeneity.
The parallel integration gives rise to the plasma dispersion function $W(%
\widetilde{\xi }_{n(s,f)\alpha })$ with the argument $%
\widetilde{\xi }_{n(s,f)\alpha }$, where the perpendicular integration yields the modified
Bessel function in the term $\Lambda _{n}(b_{(s,f)\alpha })=e^{-b_{(s,f)\alpha
}}I_{n}(b_{(s,f)\alpha })$ with the argument $\ b_{(s,f)\alpha }$.
For the static component $v_{s\alpha 0 }=0$, $\widetilde{\omega }=\omega $, and $\omega_{se}^{\ast }, \,\omega _{si}^{\ast }\neq 0$,  while  for  streaming particles $v_{f\alpha 0 }\neq 0$ and $\widetilde{\omega }=\omega -k_{z
}v_{f\alpha 0 }$, $ \omega _{f\alpha}^{\ast }=0$.
 The drift frequencies for electrons and ions are related as
$\omega _{si}^{\ast }=-\left( T_{si}/T_{se}\right)
\omega _{se}^{\ast }$ where $\omega _{se}^{\ast }>0$.

The dispersion relation for the electrostatic drift waves is obtained from linearized Poisson's equation
\[
\varepsilon_{0}k^{2}\phi _{1}=-e(n_{es1}+ n_{ef1}-n_{is1}-n_{if1}).
\]
The Larmor radii of electrons in both plasmas are very small as compared to
the ions, which allows for the expansion of the modified Bessel function for
small argument as $\Lambda _{n}(b_{(s,f) e})=[b_{(s,f) e}/2]^n/n!$.
It is easy to see that only $n=0$ terms survive in the limit of a negligible
value of the argument, i.e., for $b_{(s,f)e}\rightarrow 0$, $\Lambda
_{0}(b_{\left( s,f\right) e})=1$. We shall also separate the $n=0$ terms in
the ions contribution.

Using the identity $\Lambda _{n}(x)=\Lambda _{-n}(x)$ and  the expansion of the plasma dispersion function for $n\neq 0$ terms in
limit of the large argument, and assuming the realistic low frequency case
for both the components, i.e., $\widetilde{\xi }_{nfi},\xi _{nsi}\gg 1$ and $%
\omega \ll \Omega _{(s,f)i}$, respectively, one can easily prove that
the $n\neq 0$ terms vanish from the last terms of Eqs.~(\ref{3}-\ref{6}) and we get  the dispersion relation
\begin{equation}
\varepsilon \equiv 1+\frac{1}{k^{2}\lambda _{Dse}^{2}}\left[ A_{s}+A_{f}%
\right] =0,  \label{15}
\end{equation}%
where
\[
A_{s}=\left\{ 1+\dfrac{n_{is0}T_{se}}{n_{es0}T_{si}}+\left( 1-\frac{\omega
_{se}^{\ast }}{\omega }\right) \left( W(\xi _{0se})-1\right) \right.
\]%
\begin{equation}
\left. +\dfrac{n_{is0}T_{se}}{n_{es0}T_{si}}\left( 1-\dfrac{\omega
_{si}^{\ast }}{\omega }\right) \left[ \Lambda _{0}(b_{si})\left( W(\xi
_{0si})-1\right) \right] \right\},  \label{16}
\end{equation}%
\be
A_{f}\! =\!\frac{n_{if0}T_{se}}{n_{es0}T_{fe}}\!\left\{\!
\dfrac{T_{fe}}{T_{fi}}\!\!\left[1\!+\!\Lambda_{0}(b_{fi})\!\left( \!W(\widetilde{\xi }%
_{0fi})-1\right)\! \right] \!+\!W(\widetilde{\xi }_{0fe})\!\right\}.  \label{17}
\end{equation}%
In order to calculate the growth rate of the drift wave, we separate the real and the imaginary parts of Eq.~(\ref{15})
and the growth rate becomes
\begin{equation}
\gamma =-\frac{\varepsilon _{i}}{\partial \varepsilon _{r}/\partial \omega
_{r}}=-\frac{Im\left[ A_{s}+A_{f}\right] }{\partial (Re\left[
A_{s}+A_{f}\right] )/\partial \omega _{r}}.  \label{20}
\end{equation}%
The real dispersion relation may be obtained by taking $\varepsilon _{r}=0$,
i.e.,
\[
1+ Re\left[ A_{s}+A_{f}\right]/(k^{2}\lambda _{Dse}^{2}) =0.
\]
The wave behavior will be discussed in two different frequency limits.

\subsection{Specific frequency limits}
\paragraph{} In what follows the electrons and ions in the static component satisfy the following  frequency limits $k_zv_{tsi}\ll \omega
\ll k_zv_{tse}$, while for the streaming species we have
\be
k_zv_{tfi}\ll \left| \omega -k_{z}v_{f 0}\right|\ll k_zv_{tfe}.\label{l1}
\ee
The growth rate for the drift wave becomes
\[
\gamma_1 =-c_1(g_1+ g_2+ g_3)\equiv
-\sqrt{\frac{\pi }{2}}\dfrac{\omega _{r}^{2} n_{es0}/n_{is0}}{\omega
_{se}^{\ast }\Lambda _{0}(b_{si})}
\vspace{-0.3cm}
\]
\[
\times \left\{ \left[ \left( \frac{\omega _{r}-\omega _{se}^{\ast }}{%
k_zv_{tse}}\right) \exp \left( -\frac{\omega _{r}^{2}}{2k_{z
}^{2}v_{tse}^{2}}\right) \right. \right.
\vspace{-0.3cm}
\]
\[
\left. +\Lambda _{0}(b_{si})\dfrac{n_{is0}T_{se}}{n_{es0}T_{si}}\dfrac{%
\omega _{r}-\omega _{si}^{\ast }}{k_zv_{tsi}}\exp \left( -\frac{%
\omega _{r}^{2}}{2k_z^{2}v_{tsi}^{2}}\right) \right]
\vspace{-0.3cm}
\]
\[
+\left. \frac{n_{if0}T_{se}}{n_{es0}T_{fe}}\frac{\omega _{r}-k_{z}v_{f0}}{k_zv_{tfi}}\left[\Lambda _{0}(b_{fi})\frac{T_{fe}}{T_{fi}}\exp \!\left(\! -\frac{\left(
\omega _{r}-k_zv_{f0}\right) ^{2}}{2k_z^{2}v_{tfi}^{2}}\!\right)\!
\right.\right.
\]
\be
\left. \left.  +\sqrt{\frac{T_{fi}m_{e}}{T_{fe}m_{i}}}\exp \left( -\frac{%
\left( \omega _{r}-k_zv_{f0}\right) ^{2}}{2k_z^{2}v_{tfe}^{2}}%
\right) \right] \right\}. \label{d1}
\ee
Here, the meaning of the terms $g_1, g_2, g_3$ is obvious, and they  describe contribution of static electrons and ions, and flowing electrons and ions, respectively. The real part of the frequency may be written as
\be
\omega=\frac{\omega _{se}^{\ast }\Lambda _{0}(b_{si}) }{ \frac{n_{es0}}{n_{is0}}+ k_{\bot }^2 \rho_{ss}^2+ \frac{n_{if0}}{n_{is0}} \frac{T_{se}}{T_{fe}} \left(1+ k_{\bot }^2\rho_{sf}^2\right)}.
\label{d2}
\ee
Here,  $\rho_{s(s,f)}= c_{s(s,f)}/\Omega_i$, $c_{s(s,f)}^{2}=T_{(s,f)e}/m_i$.
The static ions term $g_2$ in Eq.~(\ref{d1})   cause  damping regardless of parameters, while  the electrons term $g_1$ yields the usual
kinetic instability provided that necessary condition $\omega_r< \omega _{se}^{\ast }$ is satisfied.

As for the contribution of the flowing
plasma, the $g_3$-term, it  turns out that the universally growing mode can completely be stabilized and this will be demonstrated below using some parameters that may be applicable to the laboratory plasma conditions.

\begin{figure}
\includegraphics[height=6cm, bb=20 15 285 217, clip=]{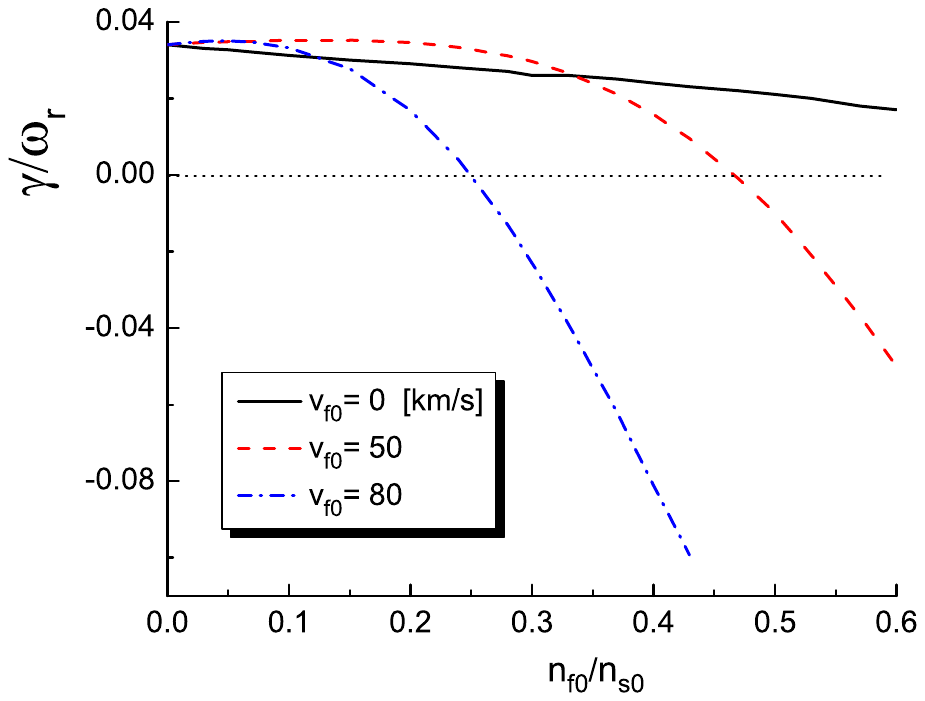}
\vspace*{-3mm} \caption{The imaginary part of the drift wave frequency  (\ref{d1}) normalized to $\omega_r$ in terms of the flowing plasma density, for the same  temperature of both plasmas.  }\label{fig1}
% \vspace{0.3cm}
\end{figure}
We choose parameters which will show that damping by $f$-plasma is  essentially due to their flow.  We take $B_0=2$ T, $L_{nsi}=L_{nse}=L_n=0.1$ m, $T_{si}=T_{se}=T_s=10^5$ K, $n_{is0}=n_{es0}=n_{s0}=10^{19}$ m$^{-3}$,  $\lambda_\bot\equiv\lambda_y=3$ mm, and take $k_z/k_y=0.0002$. For such parameters the drift wave is  unstable, $\omega_r=74659$ Hz, $\gamma_1/\omega_r=0.034$ in spite of so hot $s$-ions.
 When  $f$-plasma particles are added,  and with the same temperature $T_{fi}=T_{fe}=T_s$, there is very little change in $\gamma_1$ and the wave remains growing even if $n_{f0}$ is strongly increased, and this remains so as long as $f$-particles do not flow (see full line in Fig.~\ref{fig1}).

\begin{figure}
\includegraphics[height=6cm, bb=20 17 283 216, clip=]{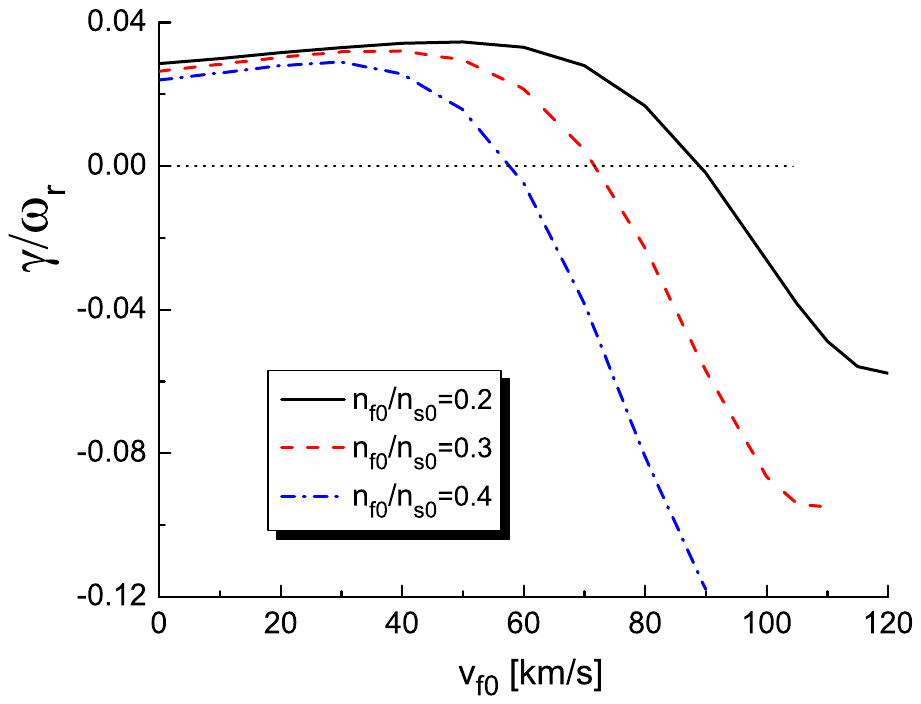}
\vspace*{-3mm} \caption{The imaginary part of the drift wave frequency  (\ref{d1}) normalized to $\omega_r$ in terms of the $f$-plasma speed, for the same  temperature of both plasmas.   }\label{fig2}
% \vspace{0.3cm}
\end{figure}

 But if the $f$-plasma is flowing,   the  growth/damping is changed. In this case  the initial instability caused by the electron term $g_1$ is first increased for small $n_{f0}$, but for larger $f$-plasma density the mode is heavily  damped.   The effect of the flow of the added plasma is thus essential,  it has a profound effect on the drift wave. See more details in Fig.~\ref{fig2} which shows that it is possible to find particular speed values for which the mode is most effectively damped. Note that here $v_{f0}<v_{ph}\equiv \omega_r/k_z=148530, 137104, 127311$ km/s for $n_{f0}/n_{s0}=0.2, 0.3, 0.4$.

\begin{figure}
\includegraphics[height=6cm, bb=18 17 276 219, clip=]{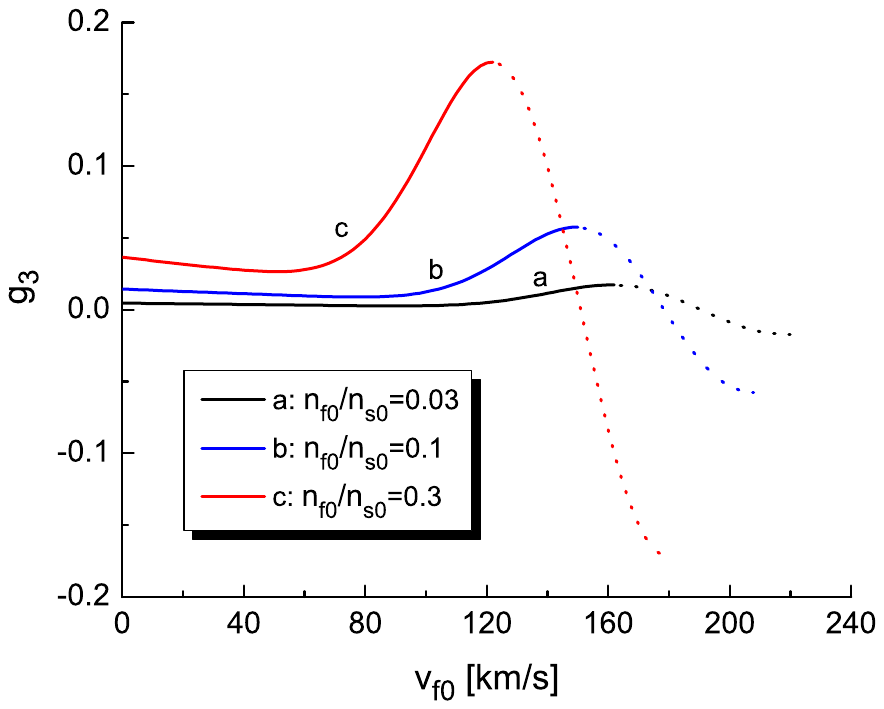}
\vspace*{-3mm} \caption{The flowing plasma term $g_3$ from Eq.~(\ref{d1}) used in Figs.~\ref{fig1},~\ref{fig2},  in terms of the $f$-plasma  speed $v_{f0}$. The  peaks are the $f$-ion exponential term contribution.   }\label{fig3}
% \vspace{0.3cm}
\end{figure}

\begin{figure}
\includegraphics[height=6cm, bb=19 17 284 219, clip=]{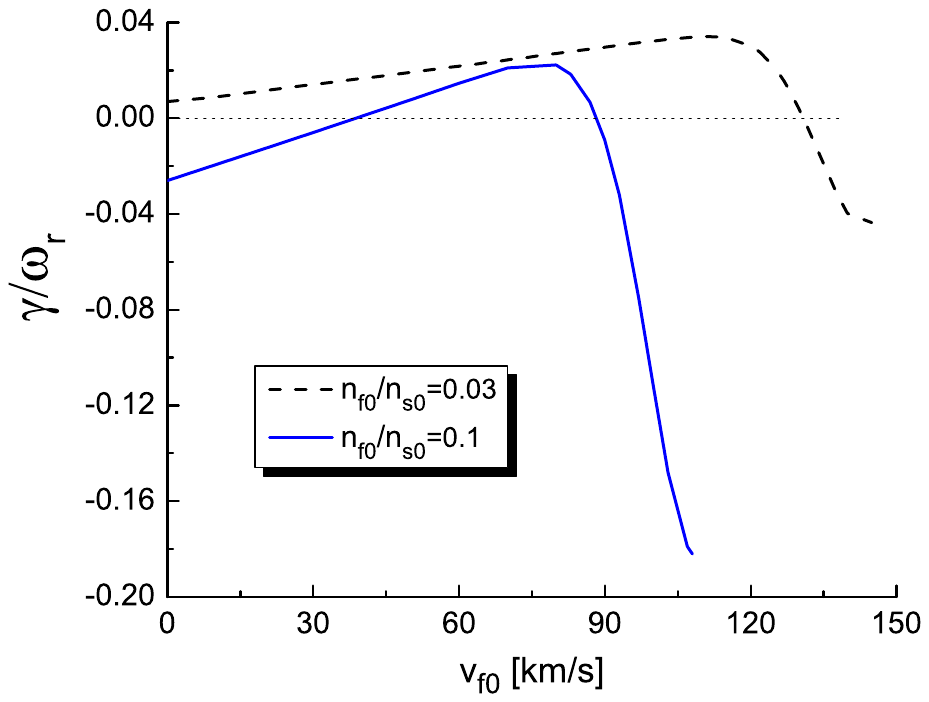}
\vspace*{-3mm} \caption{The growth rate/damping (\ref{d1}) for a cooler $f$-plasma  $T_f/T_s=1/5$ in terms of the speed $v_{f0}$.  }\label{fig4}
% \vspace{0.3cm}
\end{figure}

 The effect of the flow may be understood from Fig.~\ref{fig3}, where the $f$-plasma term $g_3$ is presented in terms of $v_{f0}$ for the three values $n_{f0}$.
 The essential part is the ion term, which is  of the shape $(v_{ph}-v_{f0})\exp[-(v_{ph}-v_{f0})^2]$, so that normally destabilizing first part $v_{ph}-v_{f0}$ is counteracted by the ion exponential part, and  the $f$-ion part in $g_3$ goes to zero for large $v_{f0}$ instead of linearly increasing the growth rate indefinitely due to $(v_{ph}-v_{f0})$ term alone (for $v_{f0}>v_{ph}$). The electrons have a  minor role  and  contribute only in the range  $v_{f0}>v_{ph}$ when the flow destabilizes the mode (not presented here). The lines are made broken to mark regions where  $k_zv_{tfi}\ll \left| \omega -k_{z}v_{f0}\right|$ is violated and this analytical model should not be used, hence the given speed limit here and in Figs.~\ref{fig4},~\ref{fig6}.

\begin{figure}
\includegraphics[height=6cm, bb=18 15 275 219, clip=]{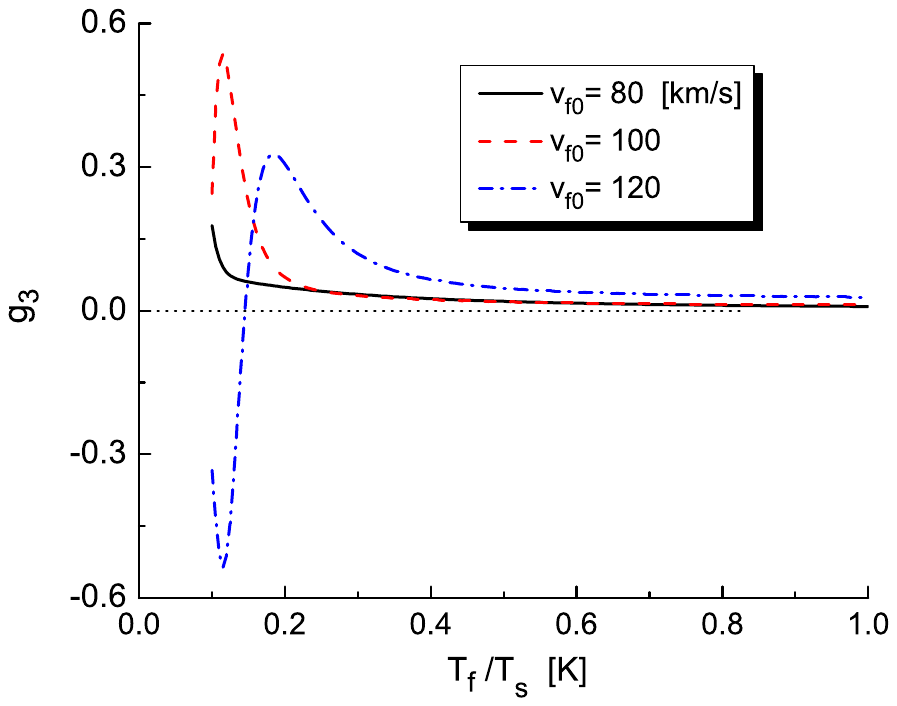}
\vspace*{-3mm} \caption{Flowing plasma term $g_3$ in the imaginary part of frequency (\ref{d1}) in terms of  $f$-plasma temperature.  }\label{fig5}
% \vspace{0.3cm}
\end{figure}

 To check the effects of the $f$-plasma temperature, in Fig.~\ref{fig4} we give the imaginary part of frequency in terms of $v_{f0}$  for the two densities $n_{f0}/n_{s0}$ and for  $T_f/T_s=0.2$.
It is seen that for $n_{f0}/n_{s0}=0.1$  the mode is immediately damped  even for $v_{f0}=0$ as soon as the $f$-plasma is added, and there is a strong damping for larger $v_{f0}$. The dependence of $\gamma_1$ on $T_f$  is complicated, but  Fig.~\ref{fig4} may partly be understood from Fig.~\ref{fig5} where  we set $n_{f0}/n_{s0}=0.1$; it is seen that in the given temperature range, around $T_f/T_s=0.2$, the $g_3$ term in (\ref{d1}) is positive and it causes strong damping, but this is not always so.

\paragraph{} In the frequency range
\be
\left|\omega -k_zv_{f0}\right| << k_zv_{tfi},
 \ee
 the $f$-ion response is nearly Boltzmannian while electron contribution is negligible (see further in the text), and with similar approximations as above we have:
\[
\gamma_{2}=-c_2(\alpha_1+ \alpha_2+\alpha_3)
\]
\[=-\sqrt{\frac{\pi }{2}}\frac{\omega _{r}^{2}}{\omega
_{se}^{\ast }\Lambda _{0}(b_{si})}  \left[
\left( \dfrac{\omega _{r}-\omega _{se}^{\ast }}{%
k_{z }v_{tse}}\right) \exp\left({-\dfrac{\omega _{r}^{2}}{2k_{z}^{2}v_{tse}^{2}}}\right)
\right.
\vspace{-0.3cm}
\]
\[
\left. +\dfrac{T_{se}}{T_{si}}\Lambda _{0}(b_{si})\left( \dfrac{\omega _{r}-\omega
_{si}^{\ast }}{k_{z }v_{tsi}}\right)\exp\left({-\dfrac{\omega _{r}^{2}}{2k_{z}^{2}v_{tsi}^{2}}}\right)
\right.
\vspace{0.3cm}
\]
\be
\left.
+\dfrac{n_{f0}T_{se}}{n_{s0}T_{fi}}\Lambda _{0}(b_{fi})\dfrac{
\omega _{r}-k_{z }v_{f0}}{k_{z }v_{tfi}}%
\right],   \label{g2}
\ee
\be
\omega_{r}\approx \frac{\omega _{se}^{\ast }}{1+ k_{\bot }^2 \rho_{ss}^2 + \frac{n_{f0}}{n_{s0}} \frac{T_{se}}{T_{fi}}}.
\label{om2}
\ee
Here, $n_{es0}=n_{es0}=n_{s0}$, $n_{ef0}=n_{if0}=n_{f0}$.  The $f$-ion term $\alpha_3$ causes a strong damping when $v_{f0}$ is small and this can be checked for the same parameters as before. However, for $v_{f0}>v_{ph}$ the wave is destabilized and this can easily be
more efficient than in the case of electron-current  driven mode \cite{hat, v1}. Indeed, in the usual electron-ion plasmas, the latter implies an additional electron current term $\alpha_4=u_0/\vte$ in the growth rate (\ref{g2}), but this can easily be smaller than the existing $\alpha_3$ term. For  $v_{f0}>v_{ph}$ we have that $\alpha_3>u_0/\vte$ if $v_{f0}/u_0> (v_{tfi}/\vte) (T_{fi}/T_{se}) (n_{s0}/n_{f0})$. Taking $\vte$ as  our $v_{tse}$, here the right-hand side can clearly be much below unity, so the ion flow in this regime can be far  more efficient in {\em exciting} the drift mode.

The omitted electron terms make only minor changes  in Eqs.~(\ref{g2},~\ref{om2}):  $\alpha_3$ term is multiplied by a small factor $1+(m_e/m_i)^{1/2}(T_{fi}/T_{fe})^{3/2}$, and the last term in denominator of Eq.~(\ref{om2}) is multiplied by a term $1+T_{fi}/T_{fe}$.

\subsection{Flowing ions case}

\begin{figure}
\includegraphics[height=6cm, bb=20 16 284 216, clip=]{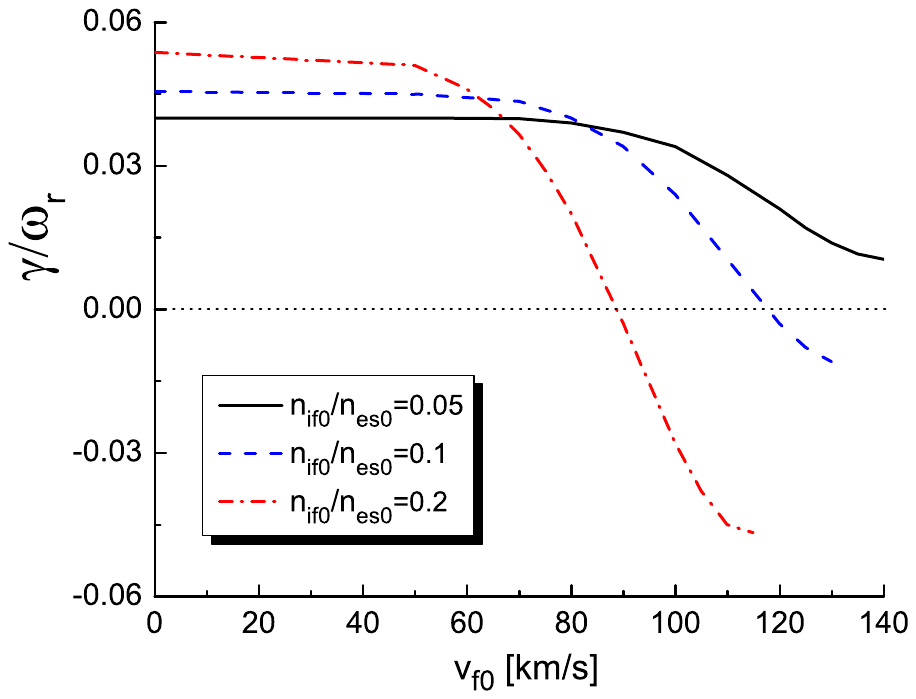}
\vspace*{-3mm} \caption{Drift wave stabilized by flowing ions.  }\label{fig6}
% \vspace{0.3cm}
\end{figure}

We checked the case of adding flowing ions only, in the range $k_zv_{tfi}\ll \left| \omega -k_{z}v_{f0}\right|$, assuming that plasma adjusts in such a way that  global quasineutrality is preserved
$n_{es0}=n_{is0} + n_{if0}$. This is completely equivalent to Ref.~\cite{has} where the stabilization is discussed by an additional cold {\em electron} population.
In Eqs.~(\ref{d1}, \ref{d2}) vanish the $f$-electron term, and the factor 1, respectively. The result is presented in Fig.~\ref{fig6} for several densities of the flowing ions and the result is similar to Fig.~\ref{fig2}. The frequency is $\omega_r=70926, 67193, 59723$ Hz for $n_{fi0}/n_{se0}=0.05, 0.1, 0.2$.  Here we keep $T_{fi}=T_{si}=T_{se}=10^5$ K, and other parameters are the same as before. The wave behavior is very similar to the previous plasma flow case.

Here, the $f$-ion flow in principle implies a current that  might cause a sheared magnetic field component $B_s=\mu_0 e n_{f0} v_{f0} L_s$, where $L_s$ is the characteristic shear length, which is known to stabilize the drift wave itself \cite{lee, wei}. However, for parameters used in the text the sheared component  is negligible; at the perpendicular distance $L_s=L_n$ it remains below $0.001 B_0$. At shorter distances it is even smaller and  can be neglected.

\section{Summary}

In conclusion, this work  provides some  clear recipes for damping of the drift wave which is usually believed to be universally unstable.  The stabilization is expected to work for  various modes from the drift wave spectrum and it can be used as an alternative for some other  mechanisms proposed in the past \cite{nis, par, kee}.

The model presented here also has an obvious advantage with respect to so called stabilization by cold electrons (having some temperature $T_c$) \cite{has} because the latter disregards simultaneous cold electron collisions with other species (which is proportional to $1/T_c^{3/2}$, so the cooler electrons the more collisions). Hence, these collisions can be
frequent even if the plasma is fairly collision-less regarding its usual components (ions and hot electrons). On the other hand, thermalization of such cold electrons is instant and its characteristic time is the same as their velocity relaxation, and they are thus totally inefficient in stabilizing the drift mode. So the ion (plasma) flow stabilization presented here is clearly a far better alternative.

\vspace{-0.5cm}
\begin{acknowledgments}
\vspace{-0.5cm}
M.F. Bashir acknowledges the Interim program for fresh PhDs by the Higher Education Commission of Pakistan.
\end{acknowledgments}

\end{document}